**Revisiting the Einstein field of a mass point**

A. LOINGER

Dipartimento di Fisica, Università di Milano

Via Celoria, 16, 20133 Milano, Italy

ABSTRACT. − The Einstein gravitational field of a material point at rest is derived anew − by a suitable limit process − from the field of a sphere of a homogeneous and incompressible fluid. This result supports clearly the thesis according to which the physically interesting singularities must correspond to the presence of matter *in loco*.

PACS. 04.20 − General relativity: Fundamental problems and general formalism − Solutions to equations.

**1.** − In the last section of his fundamental memoir on the Einstein field of a sphere of a homogeneous and incompressible fluid [1], Schwarzschild gave some hints for deriving, through a limit procedure, the Einstein field of a mass point − which had been directly deduced by him in another fundamental work [2] − from the field of the fluid. In this note, following Schwarzschild's indications, I prove that it is actually possible to deduce the field of the paper cited in [2] from the field of the paper cited in [1].

**2.** − I adopt here Schwarzschild's notations. In the following co-ordinates ($c=1$):

(2.1) $\qquad x_1 \equiv r^3/3, \quad x_2 \equiv -\cos\theta, \quad x_3 \equiv \varphi, \quad x_4 \equiv t,$

for the *internal* region of the above sphere we have (cf.[1]):

(2.2) $\qquad ds^2 = f_4\, dx_4^2 - f_1\, dx_1^2 - f_2\, \dfrac{dx_2^2}{1-x_2^2} - f_2\, dx_3^2 (1-x_2^2) \ ;$

further,

(2.3) $$f_1 f_2^2 f_4 = 1 ,$$

where the $f$'s are functions of $x_1 \equiv x$. In the external region (cf. [1] and [2])

(2.4) $$\begin{cases} (f_4)_{ext} = 1 - \alpha (3x + \rho)^{-1/3} \\ (f_2)_{ext} = (3x + \rho)^{2/3} \\ (f_1 f_2^2 f_4)_{ext} = 1 \end{cases} ,$$

The two constants of integration $\alpha$ and $\rho$ will be determined from the mass and the radius of our fluid sphere, see eqs. (33) and (34) of [1]. Schwarzschild found the functions $f(x)$ by solving Einstein equations with the following matter tensor:

(2.5) $$\begin{cases} T_1^1 = T_2^2 = T_3^3 = -p \ ; \quad T_4^4 = \rho_0 \ ; \quad T_\mu^\upsilon = 0, \text{ for } \mu \neq \upsilon \ ; \\ T := T_\alpha^\alpha = \rho_0 - 3p , \end{cases}$$

where $\rho_0$ is the invariant and constant density of the fluid, and $p=p(x)$ is the pressure [3], which must be equal to zero on the surface of the sphere (the suffix $a$ denotes, here and in the sequel, the value of a quantity on the spherical surface):

(2.6) $$p_a = p(x_a) = 0 .$$

For our aim it is not necessary to report all the results obtained by Schwarzschild in [1]. We can limit ourselves to some points.

Let us put, with our Author,

(2.7) $$f_2 = \eta^{2/3}, \quad f_4 = \zeta \eta^{-1/3}, \quad f_1 = \frac{1}{\zeta \eta} ;$$

at $x = 0$ we must have $\eta(0) = 0$, as it is easily seen. Then, externally to the sphere:

(2.8) $$\eta_{ext} = 3x + \rho, \quad \zeta_{ext} = \eta^{1/3} - \alpha ,$$

A basic clause in Schwarzschild's treatment is that the pressure must be positive and finite also at the centre of the sphere. On the contrary, to obtain the field of a mass point through a convenient "mathematical contraction" of our sphere, we must assume that at $x = 0$ both $p(x)$ and $\rho_0$ are infinite (and positive).

Since (see eqs. (10) and (22) of [1])

(2.9) $$(\rho_0 + p) \sqrt{f_4} = \text{constant} = \rho_0 \sqrt{(f_4)_a} \quad ,$$

we must determine $f_4(x)$ in such a way that $f_4(0) = 0$. Now, Schwarzschild proves (see p.430 of [1]) that for *very small* $\eta$ the function $f_4$ is given by

(2.10) $$f_4 = \frac{\lambda}{\eta^{1/3}} \left[ K + \frac{\kappa \rho_0 \sqrt{(f_4)_a}}{7} \frac{\eta^{7/6}}{\lambda^{3/2}} \right]^2 \quad ,$$

where: $(8\pi)^{-1}\kappa \equiv$ the constant $G$ of universal gravitation; $\lambda$ is an integration constant; $K$ is another constant (depending, in particular, on $\lambda$), defined by formula (27) of [1]. *If $K=0$, we have clearly $f_4(0) = 0$, as desired.* Thus

(2.11) $$f_4 = \left( \frac{\kappa \rho_0}{7 \lambda} \right)^2 (f_4)_a \, \eta^2 \quad ;$$

$\lambda$ is evidently given by

(2.12) $$\lambda = \pm \frac{\kappa}{7} \rho_0 \, \eta_a \quad ,$$

In conclusion,

(2.13) $$f_4(x) = \eta^{-2}_a (f_4)_a \, \eta^2(x) \quad ,$$

where

(2.13') $$(f_4)_a = 1 - \alpha \, (3x_a + \rho)^{-1/3} \quad .$$

If the radius $r_a$ of the sphere goes to zero we get

(2.13'') $$\lim_{x_a \to 0} (f_4)_a = 1 - \alpha \, \rho^{-1/3} \quad ,$$

in order that this limit is equal to zero we must put $\rho=\alpha^3$: but this was just the value chosen by Schwarzschild in [2], owing to a physical analogy with Newton's theory.

Next,

(2.14) $$(f_2)_a = \eta_a^{2/3} = (3x_a + \alpha^3)^{2/3} ,$$

(2.15) $$(f_1)_a = \frac{1}{(f_2)_a^2 (f_4)_a} = \frac{1}{(3x_a + \alpha^3)[(3x_a + \alpha^3)^{1/3} - \alpha]} ;$$

thus, in the limit $x_a \to 0$ we obtain actually the values at the origin $O$ of the co-ordinates for the components of the field of a mass point [2]. And the expression of $ds^2$ in the customary polar co-ordinates $\rho$, $\theta$, $\varphi$ is:

(2.16) $$ds^2 = \left(1 - \frac{\alpha}{R}\right) dt^2 - \frac{dR^2}{1 - \alpha/R} - R^2 (d\theta^2 + \sin^2\theta \, d\varphi^2) ,$$

where

(2.16′) $$R \equiv (r^3 + \alpha^3)^{1/3} ,$$

and $\alpha \equiv \kappa M/(4\pi)$ – if $M$ is the mass of the gravitating point –, as it can be easily proved, e.g., by computing the motion of a test particle at a great distance from $O$.

We have found *anew* the space-time interval corresponding to a material point *in the form of Schwarzschild's paper* [2]. And we have found **a strong argument in favour of the original choice** $\rho=\alpha^3$.

**3.** – In the literature of mathematical character on general relativity we find very many papers concerning the various kinds of singular space-times. We have, typically, quasi-regular singularities and curvature singularities, with their subspecies. From the physical standpoint, however, their interest is rather poor: as our deduction of eqs. (2.13) ÷ (2.16′) has evidenced, the physically significant singularities do correspond to a real presence of matter *in loco*, i.e. to a mass tensor different from zero. This was perfectly clear to Karl Schwarzschild – and to Albert Einstein. (In Newton's theory, the gravitational potential $U=GM/r$ is solution of Laplace-Poisson equation $\nabla^2 U(\mathbf{r}) = -4\pi\, GM\, \delta(\mathbf{r})$.)

A final remark. If you calculate the collapsing process of a spherical cloud of "dust" by relying on formulae (2.16)–(2.16′) for the external region, you will obtain a very transparent and "Galilean" result, owing to the fact that the $ds^2$ given by (2.16)–(2.16′) is quite regular for $r=\kappa\, M/(4\pi)$.

*Zum Andenken an Karl Schwarzschild* (*1873-1916*).

*———————————*